\documentclass[12pt]{article}
\usepackage{graphicx}
\usepackage{epsfig}
\newcommand{\beq}{\begin{equation}}
\newcommand{\eeq}{\end{equation}}
\newcommand{\beqa}{\begin{eqnarray}}
\newcommand{\eeqa}{\end{eqnarray}}
\newcommand{\beqar}{\begin{eqnarray*}}
\newcommand{\eeqar}{\end{eqnarray*}}

\newcommand{\labell}[1]{\label{#1}} %\qquad_{#1}} %{\label{#1}}
\newcommand{\reef}[1]{(\ref{#1})}

\newcommand{\ssc}{\scriptscriptstyle}
\newcommand{\eg}{{\it e.g.,}\ }
\newcommand{\ie}{{\it i.e.,}\ }

\newcommand{\norm}[1]{\raise.3ex\hbox{:}#1\raise.3ex\hbox{:}}

\newcommand{\al}{\alpha}

\newcommand\tp{{\widetilde p}}

\newcommand\ls{\ell_s}

\begin{document}

\setlength{\unitlength}{1mm}

\thispagestyle{empty}
\rightline{\small hep-th/0109127 \hfill } %McGill/01-xx}
\vspace*{2cm}

\begin{center}
{\bf \LARGE Superstars and Giant Gravitons}\\
\vspace*{1cm}

Robert C. Myers{$^{\dagger,}$}\footnote{E-mail: rcm@hep.physics.mcgill.ca} %;\\
%\hphantom{Om}On leave from: Department of Physics, McGill University,
%Montr\' eal, Qu\' ebec H3A 2T8, Canada.}
and
\O yvind Tafjord{$^{\spadesuit,}$}\footnote{E-mail: tafjord@wolfram.com}

\vspace*{0.2cm}

{\it $^\dagger$Perimeter Institute for Theoretical Physics}\\
{\it 35 King Street North, Waterloo, Ontario N2J 2W9, Canada}\\[.5em]

{\it $^\dagger$Department of Physics, University of Waterloo}\\
{\it Waterloo, Ontario N2L 3G1, Canada}\\[.5em]

{\it $^\dagger$Department of Physics, McGill University}\\
{\it Montr\' eal, Qu\' ebec H3A 2T8, Canada}\\[.5em]

{\it $^\spadesuit$Wolfram Research, 100 Trade Center Drive}\\
{\it Champaign, Illinois 61820, USA}\\[.5em]

\vspace{2cm} ABSTRACT
\end{center}
We examine a family of BPS solutions of ten-dimensional type IIB
supergravity. These solutions asymptotically approach AdS$_5\times
S^5$ and carry internal `angular' momentum on the five-sphere.
While a naked singularity appears at the center of the anti-de
Sitter space, we show that it has a natural physical
interpretation in terms of a collection of giant gravitons. We
calculate the distribution of giant gravitons from the dipole
field induced in the Ramond-Ramond five-form, and show that these
sources account for the entire internal momentum carried by the
BPS solutions.

\vfill \setcounter{page}{0} \setcounter{footnote}{0}
\newpage

\section{Introduction}

Brane expansion is a fascinating effect which string theory seems
to employ in a wide variety of circumstances to regulate
divergences and resolve singularities
\cite{dielectric,joestrass,giant,enhancon}. Giant gravitons
\cite{giant} provide one such example where brane expansion is
exploited in realizing the stringy exclusion principle
\cite{excluse}. Naively, type IIB supergravity in an AdS$_5\times
S^5$ background has infinite towers of BPS states associated with
excitations of the spherical harmonics on the five-sphere
\cite{fluke}. Through the AdS/CFT correspondence
\cite{adscft,adscftrev}, these seem to be related to a family of
chiral primary operators in the dual ${\cal N}=4$ U(N)
super-Yang-Mills theory. However, the latter family of operators
is cutoff because the rank of the gauge group is finite. This
cut-off is precisely reproduced by superstring theory in the
AdS$_5\times S^5$ background by realizing these states as
spherical D3-branes that expand on the background five-sphere with
increasing angular momentum \cite{giant}. The cut-off arises
because the expansion is limited by the finite volume of the
five-sphere. Hence the apparently infinite towers of BPS states
are truncated in the full string theory through the mechanism of
brane expansion.

However, the precise dictionary between the dual AdS and CFT theories
remains incomplete with regards to giant gravitons.
Recently evidence was presented that the giant gravitons
are not directly dual to single trace chiral primaries but rather
they are dual to a family of subdeterminant operators, where the finite rank
of the gauge group also imposes the same cut-off \cite{subdet}.
Further, one finds that there are `dual' giant gravitons which expand
into the AdS space and which experience no `angular momentum' cut-off
\cite{goliath,goliath2,oldref}. The test brane analysis also indicates that there
is a zero-size configuration with the same mass and internal momentum
\cite{giant,goliath,goliath2}.

So far, giant gravitons have only be analyzed as test branes
propagating in a fixed background spacetime. Certainly the AdS/CFT
dictionary would be clarified with a better understanding of
higher quantum effects on the CFT side or gravitational back
reaction on the AdS side. While the supersymmetric nature of these
states will ensure that certain properties are not modified by
such higher order corrections, we certainly expect that the
detailed dynamics of the giant gravitons will change. As a step
towards understanding gravitational back reaction, we will study a
certain family of solutions of the full nonlinear equations of
motion in type IIB supergravity. These solutions asymptotically
approach AdS$_5\times S^5$ and are charged by carrying internal
momentum on the $S^5$. Hence they have precisely the properties
one expects for solutions describing giant graviton
configurations.

To begin, we consider solutions within five-dimensional  ${\cal
N}=2$ supergravity coupled to two abelian vector supermultiplets.
This theory with a $U(1)^3$ gauge group can be embedded in the
${\cal N}=8$ theory with $SO(6)$ gauge symmetry obtained by
reducing type IIB supergravity on an $S^5$. The five-dimensional
theory has a family of black hole solutions parameterized by three
charges and a non-extremality parameter $\mu$
\cite{behrndt1,behrndt2}. The supersymmetric limit for these
solutions corresponds to $\mu=0$, however, at this point, the
event horizon has disappeared and so these BPS solutions are {\it
not} black holes. Rather they contain naked singularities, a
feature that was first noted in ref.~\cite{luke}. Given the
absence of a horizon, we refer to the BPS solutions as
`superstars'.

Now in general, one can interpret the appearance
of a naked singularity in a (super)gravity solution
as indicating the presence of an external source. In many
cases, the properties of source are such that the
solution is clearly unphysical \cite{negmass}. However,
in certain cases, one finds that the source has a
reasonable physical interpretation, especially in string theory
where the full theory has many extended brane sources --- see, \eg
\cite{source}. The main result of the present paper is to
give a physical interpretation for the source at the singularity
of the BPS solutions described above.

To address this question properly, however, we must lift the
solutions to the full ten-dimensional IIB supergravity, following
refs.~\cite{us,tenppl,duff}. We find that the BPS solutions in ten
dimensions inherit a naked singularity from their five-dimensional
reductions. In this framework, we find that a strong dipole field
is excited in the Ramond-Ramond (RR) five-form near the
singularity. Thus the source appears to be a collection of
spherical D3-branes, \ie an ensemble of giant gravitons. From the
details of the five-form field, we are able to derive the precise
distribution of giant gravitons on the five-sphere, which acts as
a source for the ten-dimensional superstars. Further we show that
the entire internal momentum of the BPS solutions is accounted for
by the giant gravitons.

An outline of the paper follows: We start in the next section by
reviewing the charged AdS$_5$ black hole solution. In Section 3, we
then present our interpretation of the singular BPS solutions in terms
of a distribution of giant gravitons. In Section 4, we present
calculations in which we probe the supergravity solutions with
dual giant gravitons. Finally, we conclude with a discussion
section.

\section{Charged AdS black holes}

To begin, we consider the $S^5$ reduction of type IIB supergravity
\cite{fluke,jhs}, truncated to the ${\cal N} =2$ sector with
$U(1)^3$ gauge symmetry. While constructing the full reduced
theory is very complex --- see, \eg \cite{vaman} --- the latter
five-dimensional theory has long been known \cite{paul} (see also,
{\it e.g.}, \cite{duff} for a review). The matter content includes
two scalar fields and the three $U(1)$ gauge fields $A_i\
(i=1,2,3)$. The scalar fields are conveniently parametrized in
terms of three scalars $X_i$ subject to the constraint $X_1 X_2
X_3=1$. This five-dimensional theory has following three-charge
black hole solutions \cite{behrndt1,behrndt2}, \beqa
ds_5^2&=&-(H_1 H_2 H_3)^{-2/3}f dt^2+(H_1 H_2
H_3)^{1/3}(f^{-1}dr^2
+r^2d\Omega_3^2),\\
X_i&=&H_i^{-1}(H_1 H_2 H_3)^{1/3},\ \ A_i=(H_i^{-1}-1)dt,
\labell{solu}
\eeqa
where
\beq
f=1-{\mu\over r^2}+{r^2\over L^2}H_1 H_2 H_3,\ \ H_i=1+{q_i\over r^2}.
\eeq
We choose the angular coordinates such that line-element on the three-sphere
is given by
\beq
d\Omega_3^2=d\al_1^2+\sin^2\!\al_1\left(d\al_2^2+\sin^2\!\al_2\,d\al_3^2\right).
\labell{threes}
\eeq
Hence the horizons appearing in these solutions have a spherical topology.
While there exist analogous solutions where the horizon has zero or constant
negative curvature, we will not discuss these solutions here. The
parameter $\mu$ is a `non-extremality' parameter, where $\mu=0$
corresponds to the supersymmetric BPS solution. The (positive) parameters $q_i$
can be related to the physical $U(1)$ charges $\tilde{q}_i$ by
\beq
\tilde{q_i}=\sqrt{q_i(\mu+q_i)}.
\eeq
For the BPS solutions, with which we will mostly be concerned,
we therefore have $\tilde{q}_i=q_i$ since $\mu=0$.
Finally the mass of this black hole is \cite{behrndt2}
\beq
M={\pi\over 4G_5}\left({3\over2}\mu+\sum q_i\right).\labell{ADMmass}
\eeq
Note that in the special case $q_1=q_2=q_3$, the scalar fields in
eq.~\reef{solu} are trivial and the solution reduces to the five-dimensional
Reissner-Nordstr\o m-anti-de Sitter geometry \cite{luke}. The thermodynamic properties
of this equal-charge solution \cite{us} and the general case \cite{cvetic,cvetic2}
were extensively studied in context of the AdS/CFT correspondence
\cite{adscft,adscftrev}.

In the absence of any scalar field excitations, the lift of the
equal-charge solution to ten dimensions is relatively
straightforward \cite{us}. The general solution \reef{solu} has
also been lifted to the ten-dimensional IIB supergravity theory
yielding the following geometry \cite{tenppl} \beqa
ds_{10}^2&=&\sqrt{\Delta}\left[-(H_1H_2H_3)^{-1}f
dt^2+(f^{-1}dr^2+r^2 d\Omega_3^2)\right]\nonumber\\
& &+{1\over\sqrt{\Delta}}\sum_{i=1}^3 H_i\left(L^2 d\mu_i^2+
\mu_i^2[L d\phi_i+(H_i^{-1}-1)dt]^2\right),
\labell{upten}
\eeqa
with
\beq
\Delta=H_1H_2H_3\sum_{i=1}^3 {\mu_i^2\over H_i},\ \
\mu_1=\cos\theta_1,\ \mu_2=\sin\theta_1\cos\theta_2,\
\mu_3=\sin\theta_1\sin\theta_2.
\labell{upten2}
\eeq
The appearance of off-diagonal terms in the above metric illustrate that the
$U(1)$ gauge charges correspond to internal momenta along the three Killing
coordinates $\phi_{1,2,3}$ in the ten-dimensional setting.
The full solution also contains a self-dual Ramond-Ramond (RR) five-form field,
which may be written $F^{(5)}=dB^{(4)}+*dB^{(4)}$ where
\beq
B^{(4)}=-{r^4\over L}\Delta\, dt\wedge d^3\Omega
-L\sum_{i=1}^3 q_i \mu_i^2(L\,d\phi_i-dt)\wedge d^3\Omega.
\labell{bfield}
\eeq
Above $d^3\Omega\equiv\sin^2\!\alpha_1\sin\alpha_2\, d^3\alpha$ denotes the volume
element on the unit three-sphere. In the limit where all the charges $q_i$
vanish, we recover the maximally supersymmetric AdS$_5\times S^5$
background with radius of curvature $L$.

We will in the following focus on the ten-dimensional BPS solution with $\mu=0$.
First we note that it is only for $\mu$ greater than some critical value that
the above solutions are true black holes with a regular horizon \cite{behrndt2}.
As a result, the BPS solutions all contain naked singularities \cite{luke}
and hence do not correspond
to extremal black holes, \ie with an event horizon with vanishing surface gravity.
This remains true whether the BPS solutions are interpreted in terms of
supergravity in ten or five dimensions. Again, given the absence of an event
horizon, we refer
the BPS solutions as `superstars'. We will briefly consider the non-BPS
configurations in the discussion section.

The naked singularities will be central in the following and so
we discuss some of their properties here. First note that the detailed
nature of the singularity
changes depending on whether only a single charge or more than one of the
charges are
nonvanishing. A common characteristic shared by all cases is
that while the singularity is located at the center of the AdS$_5$
space, it also extends over the entire $S^5$ with varying strength, \ie the rate
at which curvature invariants diverge as $r\rightarrow0$ varies with the
angles $\theta_{1,2}$.

To discuss the differences and throughout the rest of the paper,
we will assume without loss of generality that
\beq
q_1\ge q_2\ge q_3.
\eeq
If only $q_1$ is nonzero, the nonextremal horizon shrinks to zero size as $\mu$
approaches zero. At precisely $\mu=0$, the horizon disappears and $r=0$ becomes
a null singularity.
%For the BPS solution, the proper `area' of both the three-sphere in AdS$_5$ and
%the $S^5$ is vanishing at $r=0$.
If $q_1$ and $q_2$ are nonzero but $q_3=0$, the nonextremal horizon shrinks to zero
size as $\mu\rightarrow\mu_{crit}=q_1q_2/L^2$. At precisely $\mu=\mu_{crit}$,
$r=0$ becomes a null singularity. For $0\le\mu<\mu_{crit}$ (and hence
for the BPS solution), $r=0$ is a time-like singularity.
%For $\mu<\mu_{crit}$, the proper area of the three-sphere
%in AdS$_5$ remains finite, while that for the $S^5$ vanishes as $r\rightarrow0$.
Finally if all three charges are nonzero, the horizon again vanishes for $\mu$
smaller than some finite critical value. The precise value of $\mu_{crit}$
may be determined analytically \cite{behrndt2}, but here we simply observe that
$\mu_{crit}>(q_1q_2+q_1q_3+q_2q_3)/L^2$. For this solution, one can
continue the spacetime geometry past $r=0$. Defining a new radial coordinate
\beq
\rho^2=r^2+q_3,\labell{rhoeq}
\eeq
$\rho\ge0$ covers the whole geometry and the
naked singularity appears at $\rho=0$ \cite{behrndt2}. For $\mu<\mu_{crit}$,
this singularity is again timelike, and extends over the entire $S^5$ with
varying strength for generic values of the charges.
%The behavior of the spheres near $\rho=0$ depends on the details of the charges.
%The three-sphere area in AdS$_5$ remains finite if $q_{1,2}\not=q_3$ but vanishes
%otherwise as $\rho \rightarrow0$. Meanwhile the area of the $S^5$ remains finite if
%all of the charges are equal but vanishes otherwise.

\section{Superstar Goliath}

One can interpret the appearance of a naked singularity in a supergravity solution
as indicating the presence of an external source. Hence one should not necessarily
rule out such solutions as unphysical, but rather ask if it has a reasonable
physical source \cite{source}.
Here we will argue that the source appearing in the superstar solutions
has a natural interpretation within string theory as
an ensemble of giant gravitons distributed over the $S^5$
and located at the origin of AdS$_5$ geometry.

In the present context \cite{giant}, a giant graviton is a
spherical D3-brane with fixed internal `angular momentum' around a circle
of the $S^5$. Studying the equations of motion
of a test D3-brane in this configuration, one finds that it expands into a finite
sized three-sphere within the $S^5$. Given the coordinate parameterization
in eqs.~\reef{upten} and \reef{upten2}, it is convenient to consider
a giant graviton\footnote{We are only considering here the motion of
a test D3-brane in the AdS$_5\times S^5$ background, \ie in the solution
with $q_i=0$ and $\mu=0$. In the next section, we present an analysis of
the motion of certain test branes in the full superstar background.}
moving along $\phi_1$. In this case, the expansion occurs in the $\theta_1$
direction and the resulting three-sphere is parameterized by $\theta_2, \phi_2,
\phi_3$ with a (proper) area of $2\pi^2L^3\sin^3\!\theta_1$.
The calculations of ref.~\cite{giant} show that
for angular momentum $P_{\phi_1}$, the size of the giant graviton is
given by
\beq
\sin^2\theta_1 = {P_{\phi_1}/N},
\labell{sinang}
\eeq
while the energy is
\beq
E=P_{\phi_1}/L\labell{genergy}
\eeq
in agreement with the BPS bound.
In eq.~\reef{sinang}, $N$ indicates the number of
flux quanta for the RR five-form in the background. This integer
is related to the radius of curvature of the AdS$_5$ and $S^5$ geometries by
\cite{adscftrev}
\beq
L^4 = 4\pi g_s N\ls^4 %\alpha'^2.
\labell{l4eq}
\eeq
and so is implicitly understood to be large (and positive). Some other
useful relations in the following
are for Newton's constant (in five and ten dimensions)
and the D3-brane tension:
\beq
G_5=G_{10}/(\pi^3 L^5),\qquad
G_{10}=8\pi^6 g_s^2\ls^8,\qquad
T_3=1/(8\pi^3 g_s^{\vphantom 2}\ls^4).
\labell{useful}
\eeq
In any event, as $\sin^2\theta_1\le1$, eq.~\reef{sinang} yields precisely
the desired angular momentum bound: $P_{\phi_1}\le N$.

Now we turn to the physical interpretation of the naked singularity appearing
in the superstar. The source is conjectured to be a collection of giant
gravitons. These spherical branes correspond to D3-brane dipoles. Hence
while these spheres carry no net D3-brane charge,
they will locally excite the RR five-form field.
However if we consider a small (five-dimensional)
surface which encloses a portion of the sphere, this surface will carry a net five-form
flux proportional to the number of D3-branes enclosed. That is, we must choose
a surface that is the boundary of a six-dimensional `ball' which only intersects
the three-sphere of the giant graviton once at a point.
 Hence if the above conjecture is correct,
by considering an appropriate infinitesimal near the superstar singularity,
we should find a net five-form flux and further we can use this flux to determine
the local density of three-branes at the singularity.

To begin, we consider the superstar solution with one
charge excited, \ie $q_2=q_3=0$. This configuration should correspond
to a collection of giant gravitons moving
along $\phi_1$ with a certain distribution of sizes (specified
by $\theta_1$). As described above, we measure the
density of giant gravitons sitting near a certain $\theta_1$
by integrating $F^{(5)}$ over the appropriate surface. For an individual
giant graviton, we could enclose a
point on the brane at $\theta_1$ with a small five-sphere in the $r, \theta_1,$ and $\phi_1$
directions, as well as being
spanned by the $\alpha_i$. As $\phi_1$ remains a Killing
coordinate of the supergravity solution, the giant gravitons must be smeared along
this direction, as is appropriate for a momentum state. Therefore we simply integrate over
$\phi_1$, independent of $r$ and $\theta_1$. We also integrate over
$\theta_1$ with fixed $r$, which is actually appropriate as we have a distribution
of giant gravitons along the $\theta_1$ direction, rather than a single brane at a fixed
value of $\theta_1$. From eq.~(\ref{bfield}), we read off the relevant five-form component
as
\beq
F^{(5)}_{\theta_1\phi_1\alpha_1\alpha_2\alpha_3}=
2 L^2 q_1\sin\theta_1 \cos\theta_1\sin^2\alpha_1\sin\alpha_2.
\eeq
Integrating over the relevant coordinates would then give a result
proportional to $n_1$, the total number of D3-brane spheres distributed
along the $\theta_1$ direction --- with our present conventions the
result is $16\pi G_{10} T_3 n_1$. However, if we regard the surface
as being very close to the sources (\ie the radial distance is much
smaller than the scale over which the density varies), we
can find the density of branes along $\theta_1$ by simply dropping the
integral along this direction.
The density of D3-branes at a certain $\theta_1$ is then given by
\beq
{d n_1\over d\theta_1}={N\over 4\pi^3 L^4}\int
F^{(5)}_{\theta_1\phi_1\alpha_1\alpha_2\alpha_3}d\phi_1 d^3\alpha
=N{q_1\over L^2}\sin 2\theta_1.
\labell{dn1dth1}
\eeq
The total number of giant gravitons is therefore
\beq
n_1=\int_0^{\pi/2}d\theta_1\,{d n_1\over d\theta_1}=N {q_1\over L^2}.
\eeq
These calculations provide a confirmation of our conjecture as they
indicate that the source correctly excites the five-form flux
corresponding to a distribution of spherical D3-branes.

Now we further validate this result as follows:
the test brane calculations showed that a giant graviton with `radius'
given by $\theta_1$ has an
angular momentum $P_{\phi_1}=N\sin^2\theta_1$. Applying this result to
our source conjecture, the total angular momentum in
this configuration would be given by
\beq
P_{\phi_1}=\int_0^{\pi/2}d\theta_1\,{d n_1\over d\theta_1} N\sin^2\theta_1
={N^2\over2}{q_1\over L^2},
\labell{bakk}
\eeq
which precisely coincides with the total internal momentum of the superstar!
This is most easily seen using the five-dimensional BPS relation (\ref{genergy})
which fixes the total mass of the superstar in terms of the internal momentum
to be
\beq
E={P_{\phi_1}\over L} = {N^2\over 2}{q_1\over L^3}.
\eeq
To relate this result to the mass
(\ref{ADMmass}) of the superstar solution (with $q_2=q_3=\mu=0$),
we use the relations given in eqs.~(\ref{l4eq}) and \reef{useful}.
We then find that eq.~\reef{ADMmass} may be rewritten as
\beq
M={N^2\over 2}{q_1\over L^3},
\eeq
in exact agreement with the above! Hence we have an interesting
cross-check on our proposal that the source in the superstar corresponds
to a distribution of giant gravitons.

We can easily extend these calculations to the general superstar
solution with three non-zero charges. In this case, we think of the
configuration as three distinct collections of giant
gravitons, separately moving along one of the $\phi_i$ ($i=1,2,3$).
We can think of the $S^5$ as embedded in {\bf R}$^6$ coordinates
$x^{1,\ldots,6}$,
\beq
x^{2i-1}=L\mu_i\cos\phi_i,\ \ x^{2i}=L\mu_i\sin\phi_i.
\eeq
A giant graviton moving along $\phi_i$ therefore has a radius
$\rho_i=L\sqrt{1-\mu_i^2}$. In analogy with above, calculating
the density of gravitons of
a certain radius as above then involves the $F^{(5)}$ component
\beq
F^{(5)}_{\rho_i\phi_i\alpha_1\alpha_2\alpha_3}=
{d \mu_i^2\over d\rho_i}
F^{(5)}_{\mu_i^2\phi_i\alpha_1\alpha_2\alpha_3}
=2\rho_i q_i\,\sin^2\!\alpha_1\,\sin\alpha_2.
\eeq
The corresponding density of giant gravitons for each direction is
\beq
{d n_i\over d\rho_i}=2N{q_i\over L^4}\rho_i,
\labell{distribuu}
\eeq
which agrees with eq.~(\ref{dn1dth1}) for $i=1$. With this parameterization,
a giant graviton of radius $\rho_i$ carries angular momentum $P_{\phi_i}=N\rho_i^2/L^2$,
in analogy with eq.~\reef{sinang}. With this result and our graviton distributions
\reef{distribuu}, we find that the total angular momentum carried by each
set of giant gravitons is
\beq
P_{\phi_i}={N^2\over 2}{q_i\over L^2}.\labell{pqrel}
\eeq
Just as above, these results correspond to the total angular momenta
calculated for the superstar solution, and we have
complete agreement between the BPS mass of the superstar \reef{ADMmass}
and the total energy of the giant gravitons, $E=\sum
P_{\phi_i}/L$.

\section{Dual Giant Probes}

Recently string theorists have seen that certain supergravity singularities
are resolved by the expansion of branes, \eg \cite{joestrass,enhancon}.
In the present context, as well as the giant gravitons expanding on the
five-sphere, one has dual giant gravitons which expand
in the anti-de Sitter geometry \cite{goliath,goliath2,oldref}. Hence one might be tempted
to conjecture that these configurations play a role in resolving
the superstar singularity, which
appears at the origin of the AdS space.
To investigate this possibility, we examined the behavior of dual giant
graviton probes in the superstar background. Here we present a brief
account of our calculations, however, we found no evidence favoring such
an expansion in the AdS$_5$ directions.

The dual giant gravitons are spherical D3-branes,
spanning the $S^3$ in the AdS$_5$  space at a
constant value of $r$. We  let the brane orbit on the $S^5$ at constant
angles $\theta_1$ and $\theta_2$ with fixed values of the angular
momenta, $P_{\phi_i}$, conjugate to the angles $\phi_i$. The
appropriate D3-brane probe action in the superstar background is given by
\beqa
S_3&=&-T_3\int dt
d^3\alpha\left[\sqrt{-g}+B^{(4)}_{t\alpha_1\alpha_2\alpha_3}
+\sum_i\dot{\phi}_i B^{(4)}_{\phi_i\alpha_1\alpha_2\alpha_3}\right]\\
&=&{N\over L^4}\int d^4\sigma\left[
-r^3\Delta\sqrt{{f\over H_1H_2H_3}-{1\over \Delta}\sum H_i\mu_i^2
\left(L\dot{\phi}_i+{1\over H_i}-1\right)^2}
+{r^4\over L}\Delta+L\sum q_i\mu_i^2(L\dot{\phi}_i-1)\right].\nonumber
\eeqa
We fix the value of $p_i\equiv P_{\phi_i}/N$ and after some
straightforward calculations, we find the Hamiltonian
\beqa
{\cal H}(r,\mu_i)&=&{N\over L}\left[
\sqrt{\sum{\mu_i^2\over H_i}\left(1+H_1H_2H_3 {r^2\over L^2}\right)
\left[\sum{1\over H_j\mu_j^2}\left(p_j-{q_j\mu_j^2\over L^2}\right)^2
+H_1H_2H_3{r^6\over L^6}\sum{\mu_j^2\over H_j}\right]}\right.\nonumber\\
&&\left.+\sum\left(1-{1\over H_i}\right)
\left(p_i-{q_i\mu_i^2\over L^2}\right)
-H_1H_2H_3{r^4\over L^4}\sum{\mu_i^2\over H_i}+\sum{q_i\mu_i^2\over
L^2}
\vphantom{\sqrt{\left[\sum{1\over H_j\mu_j^2}\left(p_j-{q_j\mu_j^2\over L^2}\right)^2\right]}}
\right].
\eeqa
We want to minimize this energy with respect
to $r, \theta_1$ and $\theta_2$. We find that there is a non-trivial
minimum at a finite radius,
\beq
{r^2\over L^2}=\sum_{i=1}^3\left(p_i-{q_i\mu_i^2\over L^2}\right),
\eeq
when $\theta_1$ and $\theta_2$ are such that the $\mu_i$ solves the equations
\beq\label{murel}
\mu_i^2={p_i-{q_i\mu_i^2\over L^2}\over
\sum\left(p_j-{q_j\mu_j^2\over L^2}\right)}.
\eeq
These equations also imply that $L\dot{\phi}_i=1$, which is a familiar
equation from the giant graviton analysis.
The energy at this minimum is simply
\beq
{\cal H}_{\rm min}={N\over L}\sum p_i={1\over L}\sum P_{\phi_i}
\eeq
in accord with the BPS bound.
In most regions of the parameter space there is also a degenerate
minimum at $r^2=-q_3$ or $\rho=0$. The only exceptions to
this are when $q_2=q_3=0$ and $p_2$ or $p_3$ is non-zero (then ${\cal
H}(r=0)>p_1$), or when $q_2\ne 0, q_3=0, p_3\ne 0$ (then ${\cal H}(r=0)$
is infinite) --- we will return to discussing these exceptional cases in the
next section.

Let us denote the angular momentum (divided by $N$)
associated with the background
charges as $\tp_i$, such that according to eq.~(\ref{pqrel}),
\beq
\tp_i\equiv{N\over 2}{q_i\over L^2}.
\eeq
We see that in analogy with dual giant gravitons in the AdS$_5\times S^5$
vacuum, there are
(in most cases) two solutions: one zero-size graviton, and
one puffed-up dual giant. This is consistent with our interpretation in the
previous section that the source should be simply giant gravitons
distributed on the five-sphere. That is, because of the BPS properties of this system,
we are free to add additional angular momentum by introducing dual giant gravitons.
However, our test brane
analysis shows that there is no reason that the additional momentum must
be carried by probes must expanded in the AdS$_5$ directions.
In particular, as all of the $p_i$ become small, the probe
collapses onto the singularity. Hence, our brane probe analysis
provides no compelling evidence for an expansion of the superstar source
in the AdS$_5$ directions.

Let us explicitly write out the radius (in the
$r$ coordinates) of the dual giant graviton in a few simple cases:

If only one of the probe angular momenta, say $p_1$, is non-zero, we
get from eq.~(\ref{murel}) that $\mu_1=1$, and
\beq
{r^2\over L^2}=p_1-{2\over N}\tp_1,
\eeq
very similar to the results for a pure AdS$_5\times S^5$ background with $\tp_i=0$,
\ie $r^2/L^2=p_1$ in the latter case \cite{goliath,goliath2}.
There is an interesting correction (which is small in
the large $N$ limit) due to the background. One also finds that increasing
$\tp_1$ lowers the potential barrier (in the Hamiltonian)
between the zero-size graviton and the corresponding dual giant.

Another example is when the three background charges are equal,
$\tp_1=\tp_2=\tp_3=\tp$. Then we find $\mu_i^2=p_i/\sum p_j$, such
that the radius is given by
\beq
{r^2\over L^2}=\sum p_i-{2\over N}\tp.
\labell{heads}
\eeq
This again is close to the corresponding case in pure
AdS$_5\times S^5$, where one finds $\mu_i^2=p_i/\sum p_j$ as
well as the radius $r^2/L^2=\sum p_i$. In fact, in terms of the $\rho$
coordinate (\ref{rhoeq}), eq.~\reef{heads} becomes precisely $\rho^2/L^2=\sum p_i$.

We have also considered other cases, but the general analysis for the
most generic situation is very complicated and yields no interesting insights.

The center of mass of the dual giant can be thought of as
sitting at the (singular) center of the
geometry where $r^2=-q_3$ ($\rho=0$), so we might expect this to be moving
along a null trajectory. We can check this by evaluating $ds^2$ for
$L\dot{\phi}_i=1$ and $r^2=-q_3$, and we find
\beq
\left.ds^2\right|_{\rm com}=-{\mu_3\over
L^2}\sqrt{(q_1-q_3)(q_2-q_3)}.
\eeq
We see that, curiously, this is not on a null trajectory in the
generic cases. However, in many special cases we do get $ds^2=0$,
when $p_3=0$ (which gives $\mu_3=0$) or when $q_2=q_3$.

\newpage
\section{Discussion}

We studied the properties of various charged BPS supergravity
solutions of the type IIB theory in asymptotically AdS$_5\times
S^5$ spacetime. While these spacetimes contain naked
singularities, it is natural to interpret such singularities  in
terms of an external source \cite{source}. We argued that type IIB
string theory provides such a source for these solutions in the
form of an ensemble of giant gravitons. These giant gravitons were
detected by the dipole field which they excited in the RR
five-form near the singularity. Thus the naked singularities
appearing here are no worse than, \eg those in the backgrounds
representing planar D-branes. It is interesting to speculate that
on some microscopic level, the solutions should be completely
nonsingular as the sources are essentially expanded D3-branes
which have a nonsingular core \cite{igor}.

With the advent of the AdS/CFT correspondence \cite{adscft,adscftrev},
there has been
great interest in generating and studying asymptotically AdS solutions. However,
unfortunately, one finds that many of these solutions contain naked singularities
--- see, \eg \cite{sing} --- and so the question of understanding
when these singularities are physically acceptable has received some
attention \cite{joestrass,cliffsing,malnun,gubsing,korean}. Given that the superstars are
singular solutions which we are confident have a physical interpretation
in string theory, they provide an interesting test of the various
criteria which have been proposed. The proposal of ref.~\cite{malnun}
refers to the behavior of $g_{tt}$ near the singularity, where $\partial_t$
is the Killing vector generating time translations in the dual field theory.
While naively the superstar satisfies their criteria, it seems
that their proposal should be refined to account for the present solutions.
An interesting characteristic of the superstars is that in all cases,
there is an ergoregion in the vicinity of the singularity, \ie a region
in which $g_{tt}>0$! Further, generically $g_{tt}\rightarrow+\infty$ as
one approaches the singularity and so the physical reasoning of ref.~\cite{malnun}
should be reassessed. The criterion set forward in ref.~\cite{gubsing}
requires a Poincar\'e invariant background and so cannot be applied to the
present solutions. However, ref.~\cite{gubsing} also presents an interesting physical
discussion to argue that singularities appearing in the zero temperature limit
of a family of black hole solutions should be acceptable. From this point
of view, the superstar would be deemed to contain a physically acceptable singularity.

As an interesting confirmation of giant gravitons as the source in the superstars,
we combined the distribution \reef{dn1dth1} with the angular momentum-angle
relation \reef{sinang} derived with test-branes and we found that
the total angular momentum precisely matches the total angular momentum
of the solution. While an intriguing result, it is not immediately
obvious why the $(P_{\phi_1},\theta_1)$ relation derived from a test-brane
analysis should apply in a regime where the branes are generating strong
gravitational and RR fields. For the solutions with two
(or three) charges, this reasoning seems to be consistent with a description
of two (or three) independent sets of giant gravitons separately orbiting the
five-sphere.

Within the test-brane analysis, one finds for a fixed $P_{\phi_1}$ that there
are three candidate single brane configurations: a giant graviton expanded
on the five-sphere, a dual giant expanded in the anti-de Sitter geometry
and a point-like graviton. Interpretations within the dual CFT have been
suggested for the first two of these configurations. Giant gravitons are
argued to correspond to certain subdeterminant operators \cite{subdet}, while
the dual giants seem to correspond to semiclassical coherent states \cite{goliath2}.
The role of the zero-size configuration remains open, although some
evidence has been given that these states do not exist in the CFT
\cite{goliath2}. Further there is some question as to whether these
configurations actually represent distinct states or dual descriptions
of the same states, in analogy with the results of ref.~\cite{joestrass}.
A complementary possibility is that while distinct configurations, the
underlying states may mix in the quantum theory \cite{goliath,mix}.
Hence we may ask if our present calculations can shed any light on these
issues.

First our calculations indicate that all of the internal momentum
of the superstar solutions is accounted for by giant
gravitons. In particular then, there is no evidence of any zero-size gravitons
in the ensemble. While not direct evidence that such states do not
exist, this result is certainly consistent with this hypothesis.

We also considered probe calculations. Since the five-sphere at the center
of the AdS$_5$ space is singular, we did not use giant gravitons, but rather
dual giant probes. Here the original motivation had been to investigate
whether brane expansion in the AdS directions played a role in resolving
the singularity, in analogy to the enhan\c con effect \cite{enhancon}.
However, in general
there were two stable configurations: an expanded dual giant and a point-like
state, where the probe collapsed to the origin.
Thus there was no compelling evidence that the physical source should expand
into the AdS space.
Of course, we should remark that even if the sources are said to
be point-like in the AdS space, we expect that in the full string theory
their true extent is the size of the string-scale. For the present analysis,
of course, this size is negligible compared to
the macroscopic distance scales which appear in the background
supergravity solutions.

However, there were a couple of interesting exceptions to the general
result described above. For example, if $q_1\not=0$ but $q_2=q_3=0$, a dual
giant probe with $p_2\not=0$ had a single supersymmetric minimum at $r^2/L^2=p_2$
with $E_{\ssc BPS}=N\,p_2/L$. The Hamiltonian also has a local minimum at $r=0$ but
$E=N\,p_2/L\,\sqrt{1+2\tilde{p}_1/N}>E_{\ssc BPS}$.
Hence it seems that to add momentum along $\phi_2$ to the original configuration
the resulting BPS configuration must be extended in the AdS space.

Now it is interesting to note then that in the BPS background with both
$q_1\not=0\not=q_2$ (and $q_3=0$), if we examine the volume of the three-sphere
in the AdS space as $r\rightarrow0$ that the limiting volume remains finite,
\ie $V_{\Omega_3}\rightarrow2\pi^2(q_1q_2)^{3/4}|\mu_3|^{3/2}$.
This result seems consistent with the previous
observation in that the source in the full
solution of the nonlinear supergravity equations also remains extended
in the AdS space.

There is a similar result for dual probes with $p_3\not=0$ in the background with
$q_1\not=0\not=q_2$ and $q_3=0$. Here again the dual probes have
a single supersymmetric minimum with $r^2/L^2>0$. Similarly the
generic background solution with all three charges excited has the
property that the volume of the three-sphere in the AdS space has
a finite limit as $\rho\rightarrow0$.

We would interpret these results as indicating that there is a
mixing between the giant gravitons and their dual giant cousins,
at least in configurations with more than one charge excited.
No such expansion occurs for the simplest case where only $q_1$ is nonvanishing.
That is,  $V_{\Omega_3}\rightarrow0$ as $r\rightarrow0$ in this supergravity
solution, and the dual giants carrying $P_{\phi_1}$ in this background collapse
to zero size as the momentum gets small. Hence this mixing is not of
the type envisioned in refs.~\cite{goliath,mix}. Rather it results from
the interaction of different types of giant gravitons carrying internal
momentum along orthogonal directions.

We should point out that from a ten-dimensional point of view,
our superstar solutions are far from generic. They all have the distinguishing
feature that they have been lifted from solutions of the five-dimensional gauged
supergravity theory. Hence they only excite a very specific (and limited)
set of the modes in the full ten-dimensional supergravity theory.
Since the giant gravitons are all BPS and preserve a common set of supersymmetries
independent of their internal momenta \cite{goliath}, we should expect that
arbitrary distributions of giant gravitons can be arranged on the five-sphere. It might
be interesting to examine such distributions to get a better understanding
of the physics of giant gravitons. In particular, it would be interesting to
find the background solution corresponding to a single giant graviton
or rather a collection of giant gravitons, each of which carries the same $P_{\phi_1}$
and hence expands to the same size on the five-sphere.

Another property of the present solutions is that they are smeared in the
momentum directions, \ie they are invariant under rotations in $\phi_i$.
This raises two interesting
questions which deserve further investigation. First, one might argue
that this smearing is appropriate for a momentum eigenstate. On the other
hand, most of the analysis of giant gravitons regards them as classical configurations
which solve the test brane equations of motion. From this point of view, they
are inherently `localized' objects which follow a specific trajectory on the
five-sphere. It would be interesting to understand whether there is a
delocalization effect in the full quantum theory, \ie the dual field theory,
analogous to that studied in ref.~\cite{donm}. It could be that the delocalized
character of the supergravity solutions reflects an inherent property of
the giant graviton states, rather simply indicating a need for better solution
generating technology.

The smeared nature of the supergravity solutions is also interesting since
it seems this property must be related to
the `rapid' fall-off of the fields in the asymptotic AdS$_5\times S^5$ region.
Ref.~\cite{horohub} showed that the disturbance generated by a source that
is `point-like' or spherical from a ten-dimensional point of view is exponentially
large in the asymptotic region. The superstars clearly do not exhibit this
behavior, which must be because of their relatively smooth character on the five-sphere.
It would be interesting to know if such refocussing of the higher multipole moments
occurs for an individual giant graviton.
One approach to address this question would be study the gravitational
backreaction for a single giant graviton treated as a classical test brane.
A short calculation shows that the backreaction for an individual giant
graviton is small. That is, the gravitational length scale introduced would
be given by $r_g^2/L^2\simeq P_{\phi_i}/N^2\le 1/N$. Hence we expect that
this problem could be addressed within the context of linearized perturbations
of the supergravity fields in the AdS$_5\times S^5$ background.
 As the analysis of ref.~\cite{horohub} focussed on static
sources while a giant graviton is a dynamical configuration, the final result is not clear.

Studying the non-BPS solutions, \ie those with $\mu\not=0$, would also
be of interest. For the single charge case, as soon as the nonextremality
parameter is nonvanishing, an event horizon and the associated large
entropy appears immediately. In the case of more than one charge, there
is a gap between the BPS configuration and the black hole solutions.
Presumably the intervening singular but non-BPS solutions are still
physical (or perhaps only for a discrete spectrum) in analogy
with the AdS$_3$ case --- see, \eg \cite{ads3}.
In terms of the giant gravitons, one might think
these solutions are produced by adding new pairs of giant gravitons moving
in the opposite directions on the $S^5$. This will increase
the energy but not the internal angular momentum, and so supersymmetry
will be broken. Since the RR five-form gauge-field
doesn't change for non-zero $\mu$, we find the same distribution of
D3-branes as before. This can be understood from the fact that
D3-branes moving in the `wrong' direction on the $S^5$ will contribute
negative D3-brane charge.

The full story for the nonextremal configurations, and in particular
the black holes, is further complicated because it is known that
these solutions suffer from certain instabilities. This was
first suggested by the thermodynamic analysis of the charged black holes
\cite{us,cvetic,cvetic2}. Further, a fluctuation analysis
(for the analogous AdS$_4$ solutions) confirmed the existence
of a classical perturbative instability \cite{gubmitra}. It would
be interesting to understand if giant gravitons or the dual giants play
a role in the evolution of these solutions once the instability
sets in. Probe calculations may provide some insight into this question.

An obvious extension of the present work is to study the analogous
superstar solutions in eleven-dimensional supergravity. In this
case, giant gravitons appear as expanding M5- or M2-branes in
the AdS$_4\times S^7$ and AdS$_7\times S^4$ backgrounds
\cite{giant,goliath,goliath2}. Supergravity solutions describing
charged black holes have been constructed in the reduced supergravities
in four dimensions \cite{bh4} and in seven dimensions \cite{cvetic}.
Further the lift of these solutions to the full eleven-dimensional
theory has been provided in ref.~\cite{tenppl}.
Repeating the analysis of sections 3 and 4 yields very similar results
in those cases \cite{Mgiant}, and so the singularities in the corresponding superstars
are again interpreted as being generated by ensembles of giant gravitons.

\section*{Acknowledgments}

We thank Nick Evans, Clifford Johnson, Don Marolf, Simon Ross and
Matt Strassler for useful discussions.
RCM would also like to thank the Aspen Center for Physics for
hospitality during various stages of this work. This research
was supported by NSERC of Canada and Fonds FCAR du Qu\'ebec.

\end{document}